\documentclass[12pt, ams, onecolumn, amssymb, floatfix, longbibliography]{revtex4-2}

\usepackage{amsmath}
\usepackage{color}
\usepackage{amsfonts}
\usepackage{amssymb}
\usepackage{graphicx}
\usepackage{geometry}
\usepackage{hyperref}
\usepackage{comment}
\usepackage{tabularx}
\usepackage{bm}
\usepackage{euscript}
\usepackage{graphicx}
\usepackage{color}
\usepackage{amsfonts}
\usepackage{exscale}
\usepackage{amsbsy}
\usepackage{textcomp}
\usepackage{comment}
\usepackage{hyperref}
\usepackage{slashed}
\usepackage{mathtools}
\usepackage{tabularx}
\usepackage{bm}
\usepackage{euscript}
\usepackage{graphicx}
\usepackage{color}
\usepackage{amsfonts}
\usepackage{exscale}
\usepackage{amsbsy}
\usepackage{subfig}
\usepackage{textcomp}
\usepackage{comment}
\usepackage{hyperref}
\usepackage[dvipsnames]{xcolor}
\usepackage{natbib}
\usepackage{enumitem}

\usepackage{ulem}

\newcommand{\Tr}{\text{Tr}}

\numberwithin{equation}{section}

\begin{document}

\title{Current Algebra Approach to 2d Chiral Metals}
\author{Chao-Jung Lee}
\affiliation{Department of Physics, California Institute of Technology, Pasadena, CA 91125, USA}
\author{Michael Mulligan}
\affiliation{Department of Physics and Astronomy, University of California, Riverside, CA 92511, USA}

 \bigskip
 \bigskip
 \bigskip

\begin{abstract}
We reinterpret Balents and Fisher's free 2d chiral metal [Phys.~Rev.~Lett.~{\bf 76}, 2782 (1996)] as a chiral $U(N)$ Wess-Zumino-Witten model at level $k = 1$.
Here, the $U(N)$ symmetry relates the $N \rightarrow \infty$ low-energy excitations about the chiral Fermi surface.
We obtain non-Fermi liquid generalizations of the free chiral metal that maintain the $U(N)$ symmetry of the $k=1$ theory by taking the level to be a positive integer $k>1$.
We calculate two-point correlation functions of the $U(1)$ number density and current operators in these theories for general $k$.
We find $k$ to provide an overall rescaling of the amplitude of these correlation functions.
This construction illustrates the ersatz Fermi liquid proposal of Else, Thorgren, and Senthil [Phys.~Rev.~X {\bf 11}, 021005 (2021)].
\end{abstract}

\maketitle

\bigskip

\newpage

\setcounter{page}{1}

\tableofcontents

\newpage


\section{Introduction}

States of matter with a sharp Fermi surface---but whose low-energy excitations are not Landau quasiparticles---present a challenge for effective field theory.
One approach to such states begins by coupling the free Fermi gas to a gapless bosonic degree of freedom (e.g., \cite{PhysRevLett.63.680, halperinleeread, KimFurusakiWenLee1994, Altshuler1994, Polchinski1994, NayakWilczek1994short, ChakravartyNortonSyljuasen1995, OganesyanKivelsonFradkin2001, MetznerRoheAndergassen2003, AbanovChubukov2004, PhysRevB.72.045105, VARMA2002267, LawlerFradkin2007, PhysRevB.75.235116, PhysRevB.78.045109, SSLee2009OrderofLimits, MetlitskiSachdev2010Part1, Metlitski2010, Mross2010, StanfordGroup2013, StanfordGroup2014, HartnollMahajanPunkSachdev2014, KMTW2015, PhysRevLett.114.226404, PhysRevLett.123.096402, PhysRevB.92.041112, PhysRevB.92.245128, PhysRevB.92.205104, 2018ARCMP...9..227L, PhysRevResearch.2.033084, PhysRevB.98.125134, PhysRevB.103.235129, 2021arXiv210205052H}).
This boson may represent an order parameter fluctuation of the Fermi fluid in the vicinity of a quantum critical point or be an emergent gauge field in an effective description that is dual to the electron one.
If the coupling between the fermions and bosons is relevant, in the renormalization group (RG) sense, the resulting fermion + boson system generally flows towards a strongly interacting non-Fermi liquid fixed point.
Another approach---piggybacking on the generic breakdown of the Fermi liquid in $d=1$ spatial dimension---employs arrays of coupled Luttinger liquids \cite{PhysRevB.42.6623, PhysRevLett.85.2160, PhysRevLett.86.676, PhysRevB.63.054430, PhysRevB.64.045120, Plamadeala2014, PhysRevLett.124.136801}.
The resulting anisotropic states generally have power-law instabilities with nonuniversal exponents.

Here we combine this second approach with a recent proposal by Else, Thorngren, and Senthil (ETS) \cite{PhysRevX.11.021005}.
The ETS proposal is based on the IR symmetry enhancement that occurs in the Fermi liquid.
The IR symmetry is associated with the long-lived gapless quasiparticle excitations of the Fermi liquid.
The enhancement is relative to the microscopic symmetries of a free Fermi gas, such as fermion number and translation invariance.
From the effective field theory point of view \cite{shankar, Polchinski:1992ed}, the enhanced symmetry of the Fermi liquid is due to the special kinematics of the Fermi surface, which renders most quasiparticle interactions irrelevant.
A similar IR symmetry enhancement occurs  at the Luttinger liquid fixed point \cite{Fradkinbook, Giamarchibook}, for which generically no quasiparticle picture applies, and in related systems \cite{PhysRevResearch.3.023011, PhysRevB.104.014517, PhysRevB.104.235113, 2021arXiv211009492M, PhysRevB.103.165126, HUANG2021115565}.
ETS suggested the Fermi liquid symmetry enhancement may characterize a class of non-Fermi liquid metals in $d>1$, termed ersatz Fermi liquids, and showed how these IR symmetries, if preserved, constrain the properties of such states.

The ETS proposal is similar to how current algebra constrains the low-energy scattering of pions in QCD \cite{Weinberg:1996kr}.
This analogy motivates us to ask: Is there a corresponding nonlinear sigma model in which the enhanced IR symmetries of the Fermi liquid are manifest? 

In this paper, we construct one such effective theory and show how it can generalize the Fermi liquid.
We argue that the chiral Wess-Zumino-Witten (WZW) model \cite{Witten:1983ar, Polyakov:1983tt, Sonnenschein:1988ug, Stone:1989cv} with $U(N)$ symmetry at integer level $k \geq 1$ in two spacetime dimensions describes a spinless (non-)Fermi liquid metal in two spatial dimensions.
Here, $N \rightarrow \infty$ equals the number of points on the Fermi surface.
The additional spatial dimension arises from the $U(N)$ flavor degrees of freedom of the WZW model.
While we focus exclusively on 2d chiral metals, in which all excitations move in the same direction along one of the spatial dimensions, the construction appears to be generalizable non-chiral metals and/or higher dimensions.
The broken time-reversal and space inversion symmetries in the models we consider prevent all conventional low-temperature instabilities (e.g., localization or superconductivity) and therefore allows for a stable metallic state to arbitrarily low temperature.

Our construction is inspired by the seminal works of Luther \cite{Luther:1978yj}, Haldane \cite{2005cond.mat..5529H}, Castro Neto and Fradkin \cite{PhysRevB.49.10877}, and Houghton and Marston \cite{PhysRevB.48.7790} who studied the bosonization of interacting fermions in $d \geq 2$ (see also \cite{PhysRevB.52.4833}).
One difference between these earlier constructions and ours is that we use a real-space effective theory throughout.
A nonlinear bosonization scheme has recently appeared in \cite{2022arXiv220305004D}.

The remainder of this paper is organized as follows.
We begin in \S\ref{chiralmetalsection} with Balents and Fisher's free 2d chiral metal \cite{Balents96} (see also \cite{ChalkerDohmen1995, Balents97, PhysRevB.58.7619, SurLee2014}), an anisotropic free Fermi gas with half of an open Fermi surface.
This state arises from an array of $N$ coupled parallel quantum wires (a real-space analog of partitioning the Fermi surface into small nonoverlapping patches), each hosting a single chiral fermion.
We point out that this theory has a nonlocal $U(N)$ symmetry, which corresponds to transforming fermions on arbitrarily-separated wires into one another.
This observation allows us to show in \S\ref{WZWsection} that the free chiral metal is equivalent to a perturbed WZW theory with $U(N)$ symmetry at level $k=1$.
Single-fermion hopping between wires corresponds to perturbation by certain $SU(N)$ symmetry currents.
The solvability of the perturbed WZW model at $k=1$ (when it's just free fermions) extends to level $k>1$.
Since $k$ is restricted to be an integer, deformation by $k > 1$ is nonperturbative.
The resulting $k>1$ theories are not equivalent to free fermions, however, they do maintain the same symmetries as the $k=1$ theory.
It is the deformation by $k>1$ that distinguishes these models from the usual coupled Luttinger liquid constructions.
We probe these models in \S\ref{twopointcorrelators} by calculating two-point correlation functions of the $U(1)$ density and current operators for arbitrary level $k \geq 1$.
In \S\ref{discussionsection}, we conclude by discussing possible directions of future research.

\section{Nonlocal Symmetry of the Free Chiral Metal}

\label{chiralmetalsection}

In this section, we introduce the free 2d chiral metal and describe its nonlocal $U(N)$ symmetry.
A $U(1)^N$ subgroup can be identified with the momentum-space fermion densities numbering each point on the Fermi surface.

\subsection{Free Chiral Metal}

\begin{figure}
\center
\includegraphics[scale=0.5]{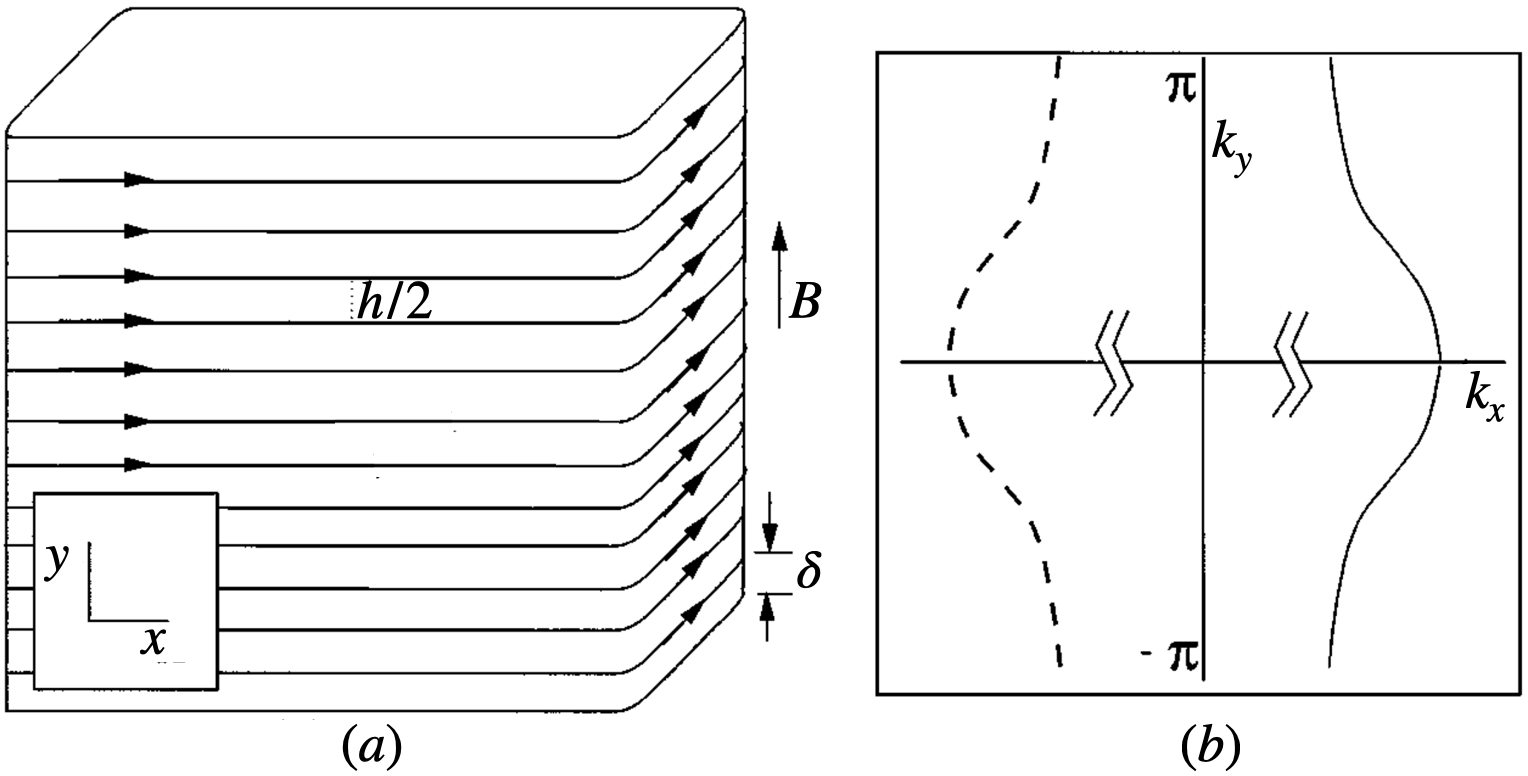}
\caption{(a) Stack of 2d integer quantum Hall states with layer separation $\delta=1$.
$B > 0$ is the strength of the magnetic field; hopping between nearest-neighbor edge modes (each of whose chirality is indicated by the counterclockwise arrow) proceeds with amplitude $h/2$.
Periodic boundary conditions along the $y$ direction are assumed.
(b) Fermi surface of the free 2d chiral metal is indicated by the solid black line. The dashed line, which would be present in a conventional system with open Fermi surface, is absent.
Both figures are slight adaptions of those in \cite{Balents96}.}
\label{chiralmetalfig}
\end{figure}

Consider a stack of $N$ integer quantum Hall states, spaced a unit distance $\delta=1$ apart from one another (see Fig.~\ref{chiralmetalfig} (a)).
In the absence of any coupling between the quantum Hall layers, the low-energy excitations of the system consist of $N$ free chiral fermion edge modes $\psi_I(x)$ with Hamiltonian,
\begin{align}
\label{S0}
H_0 = - i v  \int dx \sum_{I = 1}^N \psi^\dagger_I(x) \partial_x  \psi_I(x),
\end{align}
where the velocity $v > 0$.
The positive sign of $v$ corresponds to right-moving excitations.
The electron creation operator along the layer $I$ edge is $e^{i p_F x} \psi_I^\dagger(x)$, where $p_F^2$ is proportional to the bulk 2d electron density in each layer \footnote{For instance, in a coupled wire construction of the integer Hall state, the 2d electron density is $p_F/\pi b$, where $b$ is the wire separation.}.
The most relevant perturbation to $H_0$ consists of single-particle hopping between nearest-neighbor edges,
\begin{align}
\label{S1}
H_1 = {h \over 2} \int dx \sum_{I = 1}^N \big( \psi^\dagger_{I+1}(x) \psi_I(x) + \psi^\dagger_I(x) \psi_{I+1}(x) \big).
\end{align} 
We take the hopping amplitude $h > 0$; the overall factor of $1/2$ is for later convenience.

The total Hamiltonian $H = H_0 + H_1$ describes the free 2d chiral metal.
Taking periodic boundary conditions, $\psi_I(x) = \psi_{I+N}(x)$, along the $y$ direction and free boundary conditions along the $x$ direction, the total Hamiltonian in momentum space ${\bf k} = (k_x, k_y)$ is
\begin{align}
\label{totalmomentumH}
H = - \int d^2 {\bf k} \big(v k_x - h \cos(k_y) \big) \psi^\dagger({\bf k}) \psi({\bf k}) ,
\end{align}
where $\psi_I(x) = \int {d^2 {\bf k} \over 2\pi} e^{- i {\bf k} \cdot {\bf r}} \psi({\bf k}) \equiv {1 \over 2 \pi} \int_{-\infty}^\infty dk_x  \int_{-\pi}^{\pi} dk_y \ e^{- i {\bf k} \cdot {\bf r} } \psi({\bf k})$ and ${\bf r} = (x, I)$.
The Fermi surface is half of a conventional open Fermi surface (see Fig.~\ref{chiralmetalfig} (b)).
Using \eqref{totalmomentumH}, it consists of the $N$ points ${\bf k}_F^{(m)} = (k_x^{(m)}, k_y^{(m)})$:  
\begin{align}
k_x^{(m)} & = {h \over v} \cos\big(k_y^{(m)} \big), \\
k_y^{(m)} & = {2 \pi m \over N},
\end{align}
for $m = - {N \over 2} +1, \ldots, {N \over 2}$.

\subsection{$SU(N)$ Symmetry}

We can reinterpret the free 2d chiral metal as a 1d system of $N$ chiral fermions.
Viewing the layer $I$ label of $\psi_I$ as a flavor index, the $H_0$ part of the total Hamiltonian in \eqref{S0} is invariant under the $U(N)$ transformations:
\begin{align}
\label{freetransformation}
\psi_I \rightarrow \sum_{J=1}^N U_{IJ} \psi_J,
\end{align}
where $U_{IJ} \in U(N)$ is independent of $x$.
From the 1d perspective, this transformation is simply a $U(N)$ global symmetry.
From the 2d point of view, \eqref{freetransformation} is nonlocal since it generally relates fermions separated by an arbitrary distance $|I - J| \ \text{modulo}\ N$ along the $y$ direction.
We will refer to this transformation as a nonlocal symmetry.
There is a local $U(1)^N$ subgroup consisting of the layer $I$ phase rotations,
\begin{align}
\label{layerphaserotations}
\psi_I \rightarrow e^{i \alpha_I} \psi_I, \quad I = 1, \ldots, N,
\end{align}
where $\alpha_I$ is an arbitrary constant phase.
The hopping term in $H_1$ \eqref{S1} appears to reduce the $U(N)$ invariance to an overall $U(1)$ number symmetry.
It turns out that the full nonlocal $U(N)$ symmetry is preserved for arbitrary hopping amplitude $h$ \cite{PhysRevB.51.13449}.

To see this, we perform the gauge transformation,
\begin{align}
\label{gaugetransformation}
\psi_I(x) = \sum_{J=1}^N M_{IJ}(x) \tilde \psi_J(x),
\end{align}
with unitary $SU(N)$ matrix,
\begin{align}
\label{unitarygauge}
M(x) = \exp \Big(- i {h x \over v} \sum_{m = 1}^{N} X^{(m, m+1)}  \Big),
\end{align}
where the $SU(N)$ generators $X^{(m, n)}_{I J} = {1 \over 2} \big( \delta_{I, m} \delta_{J, n} + \delta_{I, n} \delta_{J, m} \big)$ with $m \neq n \in \{ 1, \ldots, N \}$.
As a result of the transformation \eqref{gaugetransformation}, the hopping amplitude is gauged away and the total Hamiltonian becomes
\begin{align}
\label{manifestaction}
H = - i v \int dx \sum_{I = 1}^N \tilde  \psi^\dagger_I(x) \partial_x \tilde  \psi_I(x).
\end{align}
The nonlocal $U(N)$ symmetry of $H$ is now manifest:
\begin{align}
\label{actionmanifest}
\tilde \psi_I = \sum_J U_{IJ} \tilde \psi_J \quad \leftrightarrow \quad \psi_I = \sum_{J,K,L} M_{IJ} U_{JK} M^\dagger_{KL} \psi_L.
\end{align}
Locality in the layer $I$ direction is obscured in the diagonalized Hamiltonian \eqref{manifestaction} because the fermions $\tilde \psi_I(x)$ in \eqref{gaugetransformation} are linear combinations of $\psi_I(x)$ over all $I$.
Note also that the gauge transformation alters the boundary conditions when the $x$ direction is compact. 
For example, periodic boundary conditions on a circle of length $L$ become ``twisted" by multiplication of the fermions by $M(L)$.

The charges of the nonlocal $U(N)$ symmetry are
\begin{align}
\label{noethercharges}
Q^{a} = \int dx \sum_{I, J} \tilde \psi_I^\dagger T^{a}_{IJ} \tilde \psi_J = \int dx \sum_{I, J, K, L} \psi^\dagger_I M_{I K} T^{a}_{K L} M^\dagger_{L J} \psi_J,
\end{align}
where $T^0_{IJ} = \delta_{IJ}$ and $T_{IJ}^{a}$ for $a \in \{1, \ldots, N^2 - 1 \}$ are Hermitian $SU(N)$ generators satisfying
\begin{align}
\label{SUNalgebra}
{\rm Tr} \big(T^a T^b \big) = {1 \over 2} \delta^{a b}, \quad \big[T^a, T^b\big] = i f^{abc} T^c,
\end{align} 
with $f^{abc}$ being $SU(N)$ structure constants.
The gauge transformation by $M_{IJ}(x)$ in \eqref{gaugetransformation} effects a position-dependent similarity transformation of the $U(N)$ generators,
\begin{align}
\label{similarity}
T^a_{IJ} \rightarrow  {\cal T}^a_{IJ}(x) = \sum_{K, L} M_{I K}(x) T^{a}_{K L} M^\dagger_{L J}(x),
\end{align}
under which the normalization and algebra in \eqref{SUNalgebra} are preserved.
The real-space densities defining the $Q^{a}$ are generally nonlocal with respect to the layer $I$ direction.

The conserved momentum-space densities $n({\bf k}) = \psi^\dagger({\bf k}) \psi({\bf k})$ are associated with a particular linear combination of charges $Q^a$.
To identify them, we introduce the linear combination of $SU(N)$ generators:
\begin{align}
\label{Agenerators}
A_1 & = \sum_{m=1}^N X^{(m,m+1)}, & A_2  & = \sum_{m=1}^N X^{(m, m+2)}, & \cdots\quad & , & A_{N/2} = \sum_{m=1}^N X^{(m, m+N/2)}, \\
\label{Bgenerators}
B_1 & = \sum_{m=1}^N Y^{(m,m+1)}, & B_2 & = \sum_{m=1}^N Y^{(m, m+2)}, & \cdots\quad& , & B_{N/2} = \sum_{m=1}^N Y^{(m, m+N/2)},
\end{align}
with $X^{(m,n)}$ given below \eqref{unitarygauge} and $Y^{(m,n)}_{I J} = {1 \over 2 i} \big(\delta_{I, m} \delta_{J, n} - \delta_{I, n} \delta_{J, m} \big)$ ($m \neq n$).
The $A_j$ ($B_{j'}$) matrices define nearest-neighbor, next nearest-neighbor, etc.~hopping terms of the form:
\begin{align} 
Q^{A_j} & =  \int dx \sum_{I,J} \psi_I^\dagger (A_j)_{IJ} \psi_J, \\
Q^{B_{j'}} & = \int dx \sum_{I,J} \psi_I^\dagger (B_{j'})_{IJ} \psi_J.
\end{align}
Fermions hopping via the $Q^{B_{j'}}$ set of operators acquire a $\pi/2$ phase.
Because the $A_j$ and $B_{j'}$ matrices commute with one another, the hopping terms take the same form when expressed in terms of the gauge-transformed fermions in \eqref{gaugetransformation}.
As such, these terms simply correspond to particular linear combinations of the charges $Q^a$ in \eqref{noethercharges}, the particular linear combination determined by the $SU(N)$ generators appearing in $A_j$ ($B_{j'}$).
In momentum space, these charges become
\begin{align}
Q^{A_j} & = \int d^2 {\bf k} \ 2 \cos(j k_y) n({\bf k}), \\
Q^{B_{j'}} & = \int d^2 {\bf k} \ 2 \sin(j' k_y) n({\bf k})
\end{align}
for $j, j' = 1, \ldots, N/2$.

\section{Interacting Chiral Metals}

In this section, we review the chiral WZW theory and then show how the free chiral metal can be written as a WZW theory with $U(N)$ symmetry at level $k=1$, before generalizing to arbitrary integer $k>1$.

\label{WZWsection}

\subsection{Chiral WZW Models}

The chiral WZW theory in two spacetime dimensions is a nonlinear sigma model with Wess-Zumino topological term \cite{Witten:1983ar, Polyakov:1983tt, Sonnenschein:1988ug, Stone:1989cv} (for pedagogical reviews, see \cite{2018RPPh...81d6002J, Fradkinbook}).
The WZW fixed point action for a matrix boson $g=g_{IJ}(x,t)$ taking values in a compact group $G$ is
\begin{align}
\label{chiralction}
S_k[g] = k \big( I[g] + \Gamma[g] \big),
\end{align}
where the level $k \geq 1$.
For right-moving excitations, the nonlinear sigma model term is
\begin{align}
I[g] = {1 \over 4 \pi} \int dt dx {\rm Tr} \Big( (\partial_x g^{-1} ) ( \partial_+ g) \Big),
\end{align}
where $\partial_+ = v \partial_x + \partial_t$.
(For left-movers, substitute $\partial_+ \rightarrow \partial_- = v \partial_x - \partial_t$.)
The Wess-Zumino topological term is
\begin{align}
\Gamma[g] = {1 \over 12 \pi} \int dt dx dz\ \epsilon^{\mu \nu \rho} {\rm Tr} \Big( (g^{-1} \partial_\mu g) (g^{-1} \partial_\nu g) (g^{-1} \partial_\rho g) \Big),
\end{align}
where $\mu,\nu, \rho \in \{t,x,z\}$ and the totally-antisymmetric symbol $\epsilon^{t x z} = + 1$.
In this paper, $G = U(N)$ and the trace is taken in the fundamental representation.
The Wess-Zumino term is defined on a three-dimensional hemisphere with boundary equal to the two-dimensional spacetime.
This term is independent of the extension of $g$ to three dimensions modulo $2\pi$, provided $k$ is an integer.

Under the replacement $g \rightarrow g_1 g_2$, the chiral WZW action \eqref{chiralction} satisfies the Polyakov-Wiegmann identity \cite{Polyakov:1984et}:
\begin{align}
\label{PW}
S_k[g_1 g_2] = S_k[g_1] + S_k[g_2] - {k \over 2 \pi} \int dt dx {\rm Tr} \Big(g_1^{-1} (\partial_+ g_1) (\partial_x g_2 ) g_2^{-1} \Big).
\end{align}
This identity, which is of central importance in what follows, arises from the individual multiplication rules obeyed by the nonlinear sigma model and Wess-Zumino terms:
\begin{align}
I[g_1 g_2] & = I[g_1] + I[g_2] - {1 \over 4 \pi} \int dt dx {\rm Tr} \Big( g_1^{-1} (\partial_x g_1) (\partial_+ g_2) g_2^{-1} + g_1^{-1} (\partial_+ g_1) (\partial_x g_2) g_2^{-1} \Big)
\end{align}
and
\begin{align}
\Gamma[g_1 g_2] & = \Gamma[g_1] + \Gamma[g_2] + {1 \over 4 \pi} \int dt dx {\rm Tr} \Big( g_1^{-1} (\partial_x g_1) (\partial_t g_2) g_2^{-1} - g_1^{-1} (\partial_t g_1) (\partial_x g_2) g_2^{-1} \Big).
\end{align}

The Polyakov-Wiegmann identity can be used to show that the chiral WZW theory has symmetry:
\begin{align}
\label{symmetry}
g \rightarrow V(x_-)  g W(t),
\end{align}
where $x_\pm = {1\over 2 v}(x \pm v t)$ and $V(x_-), W(t)$ are matrices in $U(N)$.
Invariance under right multiplication of $g$ by the $x$-independent matrix $W(t)$ corresponds to the fact that $g = g_R(x_-) W(t)$, with $g_R(x_-) \in U(N)$, is the general solution to the equations of motion of $S_k[g]$ that arise upon varying $g \rightarrow g + \delta g = g (1 + g^{-1} \delta g)$:
\begin{align}
\label{EOM}
\partial_x \big( g^{-1} \partial_+ g \big) = 0.
\end{align}
$g$ should therefore be thought of as a right-moving element of the coset $LU(N)/U(N)$, i.e., the loop group of $U(N)$ modulo arbitrary $x$-independent matrices $W(t)$ \cite{Stone:1989cv}. 
(Strictly speaking, the $x$ direction should be a circle, rather than the real line, for this to be the usual loop group.)
Invariance under left multiplication of $g$ by $V(x_-)$ corresponds to a right-moving $U(N)$ Kac-Moody symmetry \cite{goddard1986kac}.
Conservation of the Kac-Moody currents,
\begin{align}
\label{currents}
J^a = {i k \over 2 \pi} {\rm Tr} \Big( T^a (\partial_x g) g^{-1} \Big),
\end{align}
i.e., $\partial_+ J^a = 0$, follows upon applying \eqref{PW} to the variation, $g \rightarrow g + \delta g = \big(1 + (\delta g) g^{-1}\big) g$.
Note that $T^0_{IJ} = \delta_{IJ}$ and $T^a$ for $a = 1, \ldots N^2-1$ are the same $SU(N)$ generators appearing in \eqref{SUNalgebra}.
These Kac-Moody currents obey the usual equal-time commutation relations:
\begin{align}
\big[J^0(x, t), J^0(x', t) \big] & = - {i kN \over 2 \pi} \partial_x \delta(x - x'), \\
\big[J^a(x, t), J^b(x', t) \big] & = i f^{abc} J^c \delta(x - x') - {i k \delta^{ab} \over 4 \pi} \partial_x \delta(x - x'),
\end{align}
for $a,b = 1, \ldots N^2-1$.

When the level $k = 1$, $S_k[g]$ is equivalent to $N$ right-moving fermions with action,
\begin{align}
S_{k=1}[g] \leftrightarrow i \int dt dx \sum_{I = 1}^N \psi_I^\dagger \partial_+ \psi_I,
\end{align}
where ``$\leftrightarrow$" indicates ``equivalent to."
This equivalence is the chiral version of Witten's non-Abelian bosonization \cite{Witten:1983ar, Polyakov:1983tt}.
The Kac-Moody currents correspond to the fermion bilinears,
\begin{align}
\label{freefermioncurrents}
J^a \leftrightarrow \sum_{I,J} \psi^\dagger_I T^a_{IJ} \psi_J.
\end{align}

The level $k>1$ theories are interacting generalizations of the $k=1$ theory, in which the $U(N)$ symmetry is maintained.
One way to think about them, which will not be directly used in the remainder, goes as follows \cite{tsvelik2007quantum}.
Consider the theory of $kN$ nonchiral free Dirac fermions.
Similar to the free chiral metal, this theory arises in the low-energy limit at the surface of a stack of $kN$ integer spin-quantum Hall states.
Factorizing the nonlocal symmetry as $U(kN)_1 \approx U(1)_{kN} \times SU(N)_k \times SU(k)_N$ for each chirality, we may couple right-moving and left-moving $SU(k)_N$ currents $I^a_{R,L}$ \footnote{These currents are defined in terms of fermion bilinears as follows: $I_{R,L}^a = \sum_{I, J} (\psi^\dagger_{R,L})_I \big( T^a \otimes \mathbb{I} \big)_{IJ} (\psi_{R,L})_J$, where $T^a$ generate an $SU(k)$ subgroup of $SU(k N)$ and $\mathbb{I}$ is the identity matrix in the complementary $SU(N)$ subgroup.} through the marginally-relevant interaction,
\begin{align}
\label{conformalembeddinginteraction}
S' = - c \int dt dx \sum_{a = 1}^{k^2-1} I^a_R I^a_L.
\end{align}
For $c > 0$, the interaction \eqref{conformalembeddinginteraction} drives the free fermion system to the nonchiral $U(N)_k$ WZW model \cite{tsvelik19871+}. 
Although the interaction \eqref{conformalembeddinginteraction} is nonlocal in the flavor coordinates from the perspective of the underlying fermions, the resulting fixed point theory, i.e., the $U(N)_k$ WZW model with $k>1$, has the same symmetry as the $k=1$ theory.
$S_k[g]$ in \eqref{chiralction} is the chiral version of this fixed point in which there are only right-moving excitations.

\subsection{Perturbed WZW Models}

To make contact with the free chiral metal, we need to add nearest-neighbor hopping, i.e., the $H_1$ term in \eqref{S1}, to the $k=1$ WZW model in \eqref{chiralction}.
Using the identification of $U(N)$ currents in \eqref{freefermioncurrents}, we see that the free chiral metal is equivalent to 
\begin{align}
\label{WZWfreechiralmetal}
S_{k=1}[g] - {h \over 2 \pi} \int dt dx {\rm Tr} \Big( A_1 (\partial_x g) g^{-1} \Big),
\end{align}
where $A_1$ is the linear combination of $SU(N)$ generators given in \eqref{Agenerators}.
The second term in \eqref{WZWfreechiralmetal} corresponds to minimal coupling to a constant vector potential polarized in the $A_1$ direction of the $SU(N)$ group.
As in the free fermion representation of this fixed point, we can gauge away the hopping term to keep the $U(N)$ Kac-Moody symmetry manifest. 
If we replace $g \rightarrow M g$ with $M$ given in \eqref{unitarygauge}, the Polyakov-Wiegmann identity \eqref{PW} gives
\begin{align}
S_{k=1}[M g] - S_{k=1}[M] = S_{k=1}[g] - {h \over 2 \pi} \int dt dx {\rm Tr} \Big( A_1 (\partial_x g) g^{-1} \Big).
\end{align}
Because $M$ is a gauge transformation parameter, i.e., nondynamical, we can absorb $\exp(-i S_{k=1}[M])$ into the overall normalization of the path integral for the WZW model.
The final step is to define the gauge-transformed bosons as $\tilde g = M g$. 
The path integration measure is invariant with respect to $g \rightarrow M g$.

All of these manipulations carry over for general integer $k > 1$.
The resulting interacting chiral metals have the same symmetries as the free chiral metal.
In analogy with the $k=1$ theory, we take the $k>1$ chiral metals to be perturbed WZW models:
\begin{align}
{\cal S}_k[g] = S_{k}[g] - {h k \over 2 \pi} \int d^2x {\rm Tr} \Big( A_1 (\partial_x g) g^{-1} \Big).
\end{align}
The manifestly $U(N)$-invariant form obtains by taking $\tilde g = M g$ with action, $S_k[\tilde g]$, and $M$ given in \eqref{unitarygauge}.
The $U(N)$ symmetry currents in this ``tilded basis" are
\begin{align}
\tilde J^a = {i k \over 2 \pi} {\rm Tr} \Big(T^a (\partial \tilde g) \tilde g^{-1} \Big).
\end{align}
In the next section, we will make use of the standard equal-time two-point correlation functions of these currents:
\begin{align}
\label{correlatorcharge}
\big\langle \tilde J^0(x) \tilde J^0(x') \big\rangle_k & = \big({i \over 2 \pi} \big)^2 {N k \over (x - x')^2}, \\
\label{correlatorSUN}
\big\langle \tilde J^a(x) \tilde J^b(x') \big\rangle_k & = \big({i \over 2 \pi} \big)^2 {k \over (x - x')^2} {\delta^{ab} \over 2}, \quad a,b \in \{1, \ldots N^2 - 1 \},
\end{align}
with all other two-point correlation functions equal to zero, where $\tilde J^0 = J^0$ is the overall $U(1)$ current associated with the generator $T^0_{IJ} = \delta_{IJ}$.

\section{Two-Point Correlation Functions}

\label{twopointcorrelators}

In this section, we calculate two-point correlation functions of the $U(1)$ number density and $U(1)$ current along the $y$ direction, i.e., along the flavor space direction, in chiral metals at arbitrary interaction $k \geq 1$.

\subsection{The Density Correlation Function}

In a free Fermi gas, the equal-time $U(1)$ number density two-point function (a.k.a.~the pair correlation function) gives the relative probability of finding two particles at ${\bf r}$ and ${\bf r}'$ \cite{baym2018lectures}.
The vanishing of this correlation function at ${\bf r} = {\bf r}'$ is a reflection of the Pauli exclusion principle.
The finite, nonzero fermion density produces a nonzero asymptote as $|{\bf r} - {\bf r}'| \rightarrow \infty$.
Oscillations in the correlation function are determined by the Fermi wave vector.

We are interested in computing the analog of this correlation function in the chiral metals with action ${\cal S}_k[g]$ for general $k \geq 1$.
We focus on the long-distance, connected part of this two-point correlation function.
This means that we will not directly probe the quantum statistics of the underlying interacting particles (since the insertion points will never be coincident) and that the correlator will vanish as $|{\bf r} - {\bf r}'| \rightarrow \infty$. 
As may be anticipated from the expression in \eqref{currents} for the symmetry currents, we will find that the density correlation function for $k>1$ coincides with the $k=1$ result, up to an overall factor of $k$.
Thus, the ``rate" or amplitude at which this correlation function vanishes as $|{\bf r} - {\bf r}'| \rightarrow \infty$ gives a measure of the interaction parameterized by $k$.
A similar conclusion obtains from the current two-point function studied in the next section.

To begin, we define the local $U(1)$ number density at ${\bf r} = (x,I)$ as
\begin{align}
\rho_I(x) = \sum_a^I c_a J^a(x) = {i k \over 2 \pi} {\rm Tr} \Big( \sum_a^I c_a T^a \big(\partial_x g(x)\big) g^{-1}(x) \Big).
\end{align}
In this and future expressions, the time dependence is left implicit with all operators evaluated at the same time; the coefficients $\{ c_a \}$ in the above sum of $U(N)$ generators are chosen so that $\sum_a^I c_a T_{KL}^a = \delta_{IK} \delta_{IL}$ (no sum over $I$) \footnote{For example, when $N = 2$, $\delta_{1K} \delta_{1L} = {1 \over 2} \mathbb{I}_{KL} + {1 \over 2} (\sigma_3)_{KL}$ and $\delta_{2K} \delta_{2L} = {1 \over 2} \mathbb{I}_{KL} - {1 \over 2} (\sigma_3)_{KL}$, where $\mathbb{I}$ is the $2 \times 2$ identity matrix and $\sigma_3$ is the usual Pauli matrix.}.
When $k=1$, $\rho_I(x)$ is the bosonized expression for the fermion bilinear $\psi^\dagger_I \psi_I(x)$ (no sum over $I$).
We are interested in computing the two-point density correlation function,
\begin{align}
\langle \rho_I(x) \rho_K(x') \rangle_k
\end{align}
for nonzero hopping $h$ at $k \geq 1$.
(When $h = 0$, the $\langle \rho_I(x) \rho_K(x') \rangle_k \propto \delta_{IK}/|x-x'|^2$.)

We first factor out the $k$ dependence in order to relate the $k>1$ correlation functions to the $k=1$ correlation function:
\begin{align}
\label{generalktoonedensity}
\langle \rho_I(x) \rho_K(x') \rangle_k = k \langle \rho_I(x) \rho_K(x') \rangle_{k=1}.
\end{align}
To do this, we switch to the ``tilded basis," $\tilde g = M g$ with $M$ given in \eqref{unitarygauge}.
Up to an overall additive constant (that we ignore), the density becomes
\begin{align}
\rho_I(x) = {i k \over 2 \pi} {\rm Tr} \Big(\sum_a^I c_a M T^a M^\dagger \tilde g \partial_x \tilde g \Big).
\end{align}
We may decompose $M T^a M^\dagger$ in terms of the $U(N)$ generators as
\begin{align}
M(x) T^a M^\dagger(x) = \sum_b \beta^a_b(x) T^b,
\end{align}
for some expansion ``coefficients" $\{ \beta_a(x) \}$.
The density correlation function becomes
\begin{align}
\langle \rho_I(x) \rho_K(x') \rangle_k & = \sum_{a}^I \sum_b^K \sum_m \sum_n c_a c_b \beta^a_m(x) \beta^b_n(x') \big \langle \tilde J^m(x) \tilde J^n(x') \big \rangle_{k}.
\end{align}
Notice that the only dependence on $k$ occurs in the current two-point functions.
Replacing $\big \langle \tilde J^m(x) \tilde J^n(x') \big \rangle_{k} \rightarrow k \big \langle \tilde J^m(x) \tilde J^n(x') \big \rangle_{k=1}$ using \eqref{correlatorcharge} and \eqref{correlatorSUN}, we may invert the above relations in the $k=1$ theory to obtain the desired result in \eqref{generalktoonedensity}.

\begin{figure}
\center
\includegraphics[scale=0.8]{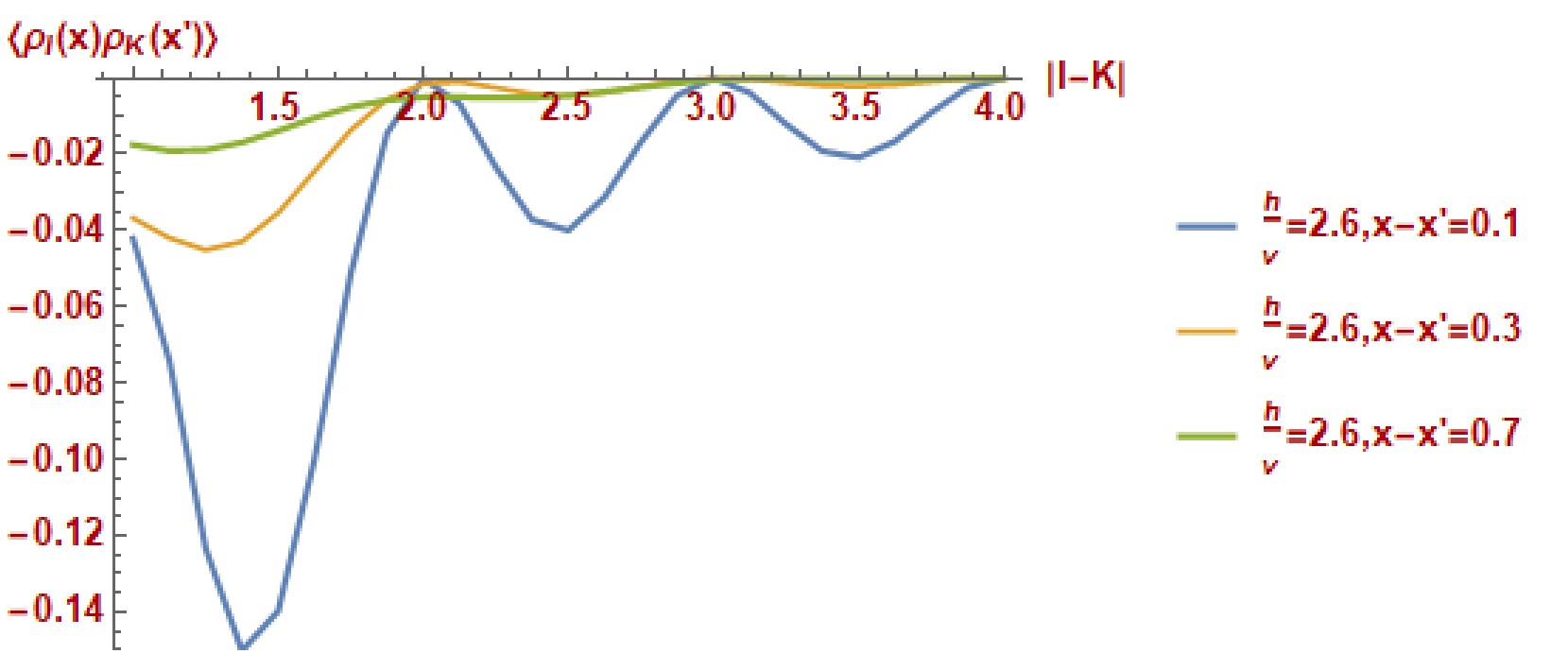}
 \includegraphics[scale=0.8]{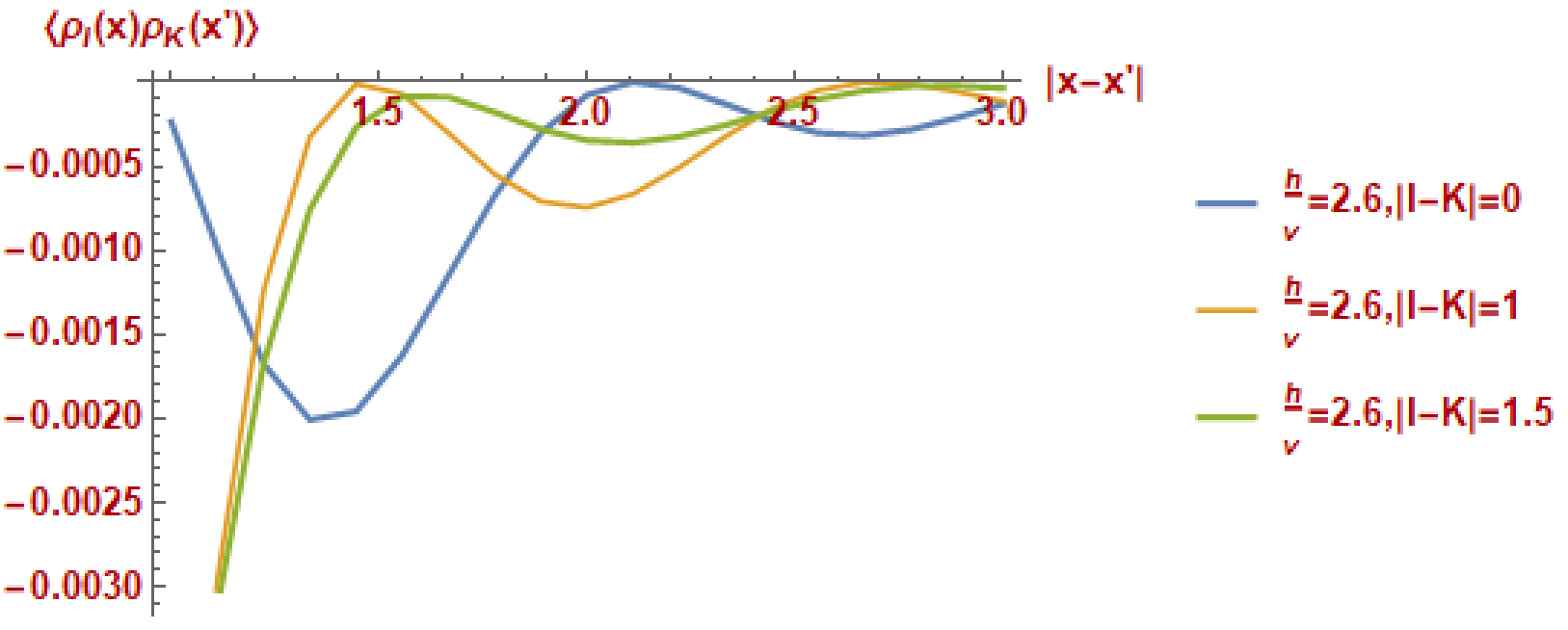}
\caption{Density two-point functions, normalized by $k$, as a function of $|I-J|$ at fixed $|x - x'|$ (top) and $|x - x'|$ at fixed $|I - K|$ (bottom).}
\label{xandydirectionsdensity}
\end{figure}

\begin{figure}
\center
\includegraphics[scale=0.8]{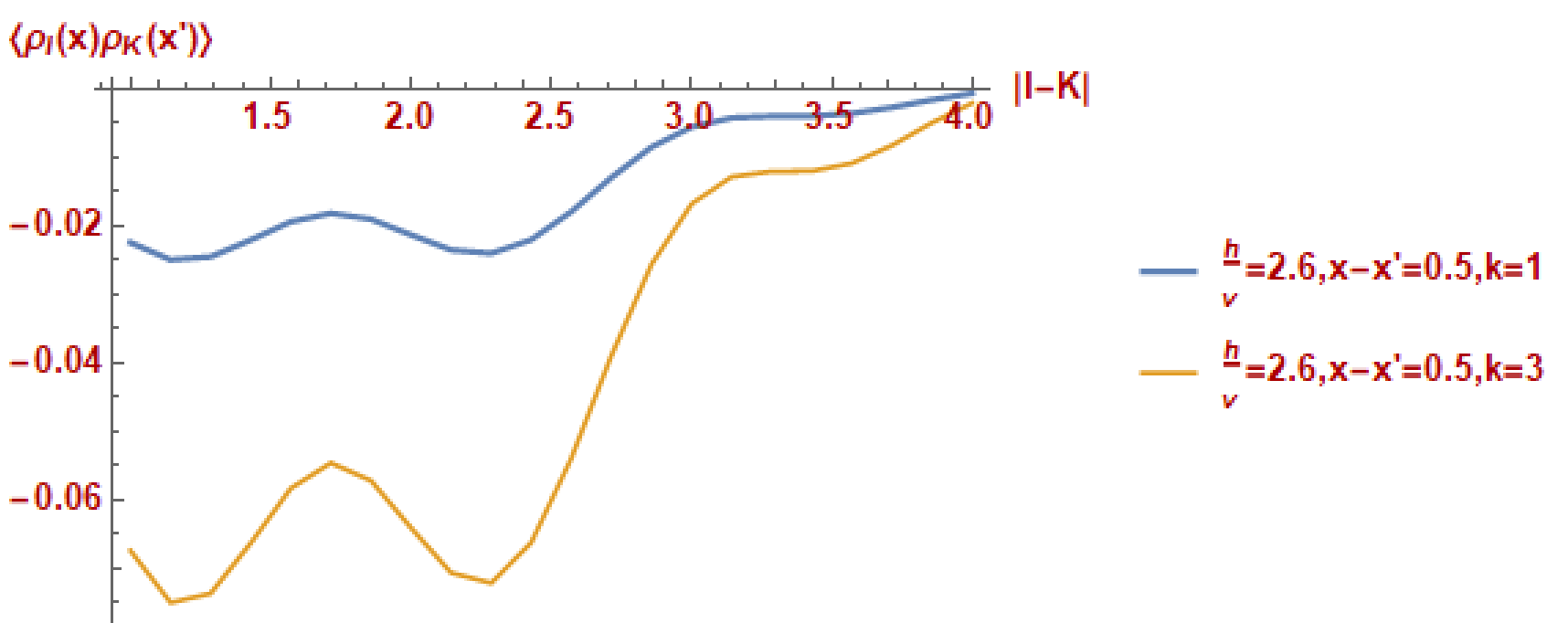}
\caption{Comparison of the density two-point functions at $k = 1$ and $k=3$ as a function of $|I-J|$ at fixed $|x - x'|$.}
\label{levelkcomparison}
\end{figure}

We now compute the $k=1$ density two-point function directly using the free fermion representation of the theory: 
\begin{align}
\langle \rho_I(x) \rho_K(x') \rangle_{k=1} = \langle : \psi^\dagger_I \psi_I(x) \psi^\dagger_K \psi_K(x') : \rangle.
\end{align}
The colons denote normal ordering, which in practice here means that we compute the connected part of this fermion four-point function.
Going to the  ``tilded basis" \eqref{gaugetransformation}, we encounter
\begin{align}
\label{intermediatestep}
\langle \rho_I(x) \rho_K(x') \rangle_{k=1} = - {1 \over 4 \pi^2} \Big(M(x) M^\dagger(x') \Big)_{IK} \times \Big(M(x') M^\dagger(x) \Big)_{KI} {1 \over (x - x')^2},
\end{align}
where we used the ``tilded basis" fermion two-point functions, 
\begin{align}
\langle \tilde \psi_I(x) \tilde \psi^\dagger_K(x') \rangle = {i \over 2 \pi} {\delta_{I K} \over x - x'}.
\end{align}
We will not need to determine the explicit form of the expansion coefficients $\{ c_a \}$ or $\{ \beta^a_b \}$ that occur above. 
Instead we evaluate the product of matrix elements by diagonalizing the matrix $A_1$, which occurs in $M(x)$ (see \eqref{unitarygauge} and \eqref{Agenerators}):
\begin{align}
A_1 = U \Lambda U^\dagger \equiv \begin{pmatrix}
u_1 & u_2 & \ldots u_N
\end{pmatrix}
\begin{pmatrix}
\lambda_1 & 0 & 0 .. \\
  0&  \lambda_2& 0 .. \\
  0 & 0 &   \lambda_3 ..\\
  ..
\end{pmatrix}
\begin{pmatrix}
u_1 & u_2 & \ldots u_N
\end{pmatrix}^\dagger,
\end{align}
where
\begin{align}
u_l = \sqrt{ \frac{2}{N }  }
\begin{pmatrix}
\sin\Big(\frac{2\pi l \times 1}{N }  -\frac{\pi}{4} \Big)\\
\sin\Big(\frac{2\pi l \times 2}{N } -\frac{\pi}{4}  \Big)\\
\vdots \\
\sin\Big(\frac{2\pi l \times N}{N }  -\frac{\pi}{4}   \Big)\\
\end{pmatrix}  
, \;
\lambda_l=  2   \,  \cos\Big(\frac{2\pi l }{ N }  \Big) 
, \; l=1,2, \ldots, N,
\end{align}
and substitute into \eqref{intermediatestep}.
The product of matrix elements takes a functional form that allows us to evaluate $|I-K|$ for noninteger values. 
We plot the density two-point functions when $N = 40$ in Figs.~\ref{xandydirectionsdensity} and \ref{levelkcomparison}.
This value of $N$ appears to be sufficiently large to accurately capture the $N \rightarrow \infty$ limit.
The oscillations in these curves are due to the Fermi surface.
We comment that $j$th-neighbor hopping may be included similarly, since the various $A_j$ and $B_{j'}$ matrices (appearing in \eqref{Agenerators} and \eqref{Bgenerators}) matrices commute with one another.

\subsection{The Current Correlation Function}

The $U(1)$ current density along the $x$ direction is equal to the (charge) density $\rho_I(x)$. 
Its two-point function therefore coincides with the $\rho_I(x)$ two-point function studied in the previous section.
In this section, we therefore focus on the two-point correlation function of the $U(1)$ current along the $y$ direction.

It is simplest to define the $U(1)$ current by way of the free fermion representation of the $k=1$ WZW theory.
To this end, we use the Peierls substitution to introduce a gauge field $A^y$ polarized along the $y$ direction into the hopping term $H_1$ in \eqref{S1}:
\begin{align}
H_1[A^y] = {h \over 2} \int dx \sum_{I, J} \delta_{J, I+1} \psi^\dagger_I e^{i A^y_{IJ}(x)} \psi_J + {\rm h.c.}
\end{align} 
To linear approximation in $A^y$, the corresponding current (density) along the $y$ direction is
\begin{align}
J_I^y(x) & \equiv \Big({\delta H_1 \over \delta A^y_{I, I+1}} + {\delta H_1 \over \delta A^y_{I-1, I}} \Big)\Big|_{A^y = 0} \cr
& = h \sum_{K, L} \psi_K^\dagger P^I_{KL} \psi_L,
\end{align}
where the matrix,
\begin{align}
P^I_{KL} = {1 \over 2 i} \Big( \delta_{K, I-1} \delta_{L, I} - \delta_{K, I} \delta_{L, I -1} + \delta_{K, I} \delta_{L, I+1} - \delta_{K, I+1} \delta_{L, I} \Big).
\end{align}
$P^I$ coincides with a linear combination of the $U(N)$ generators $Y^{(I,I+1)}$ and $Y^{(I-1, I)}$ defined below \eqref{Bgenerators}. 
For general $k \geq 1$, we therefore take the $U(1)$ current densities along the $y$ direction to be
\begin{align}
J^y_I(x) = {i k \over 2 \pi} {\rm Tr} \Big(P^I \big(\partial_x g(x)\big) g^{-1}(x) \Big).
\end{align}
Having defined the $y$ current $J^y_I(x)$, we set $A^y=0$ in the remainder.

\begin{figure}[h]
\center
\includegraphics[scale=0.8]{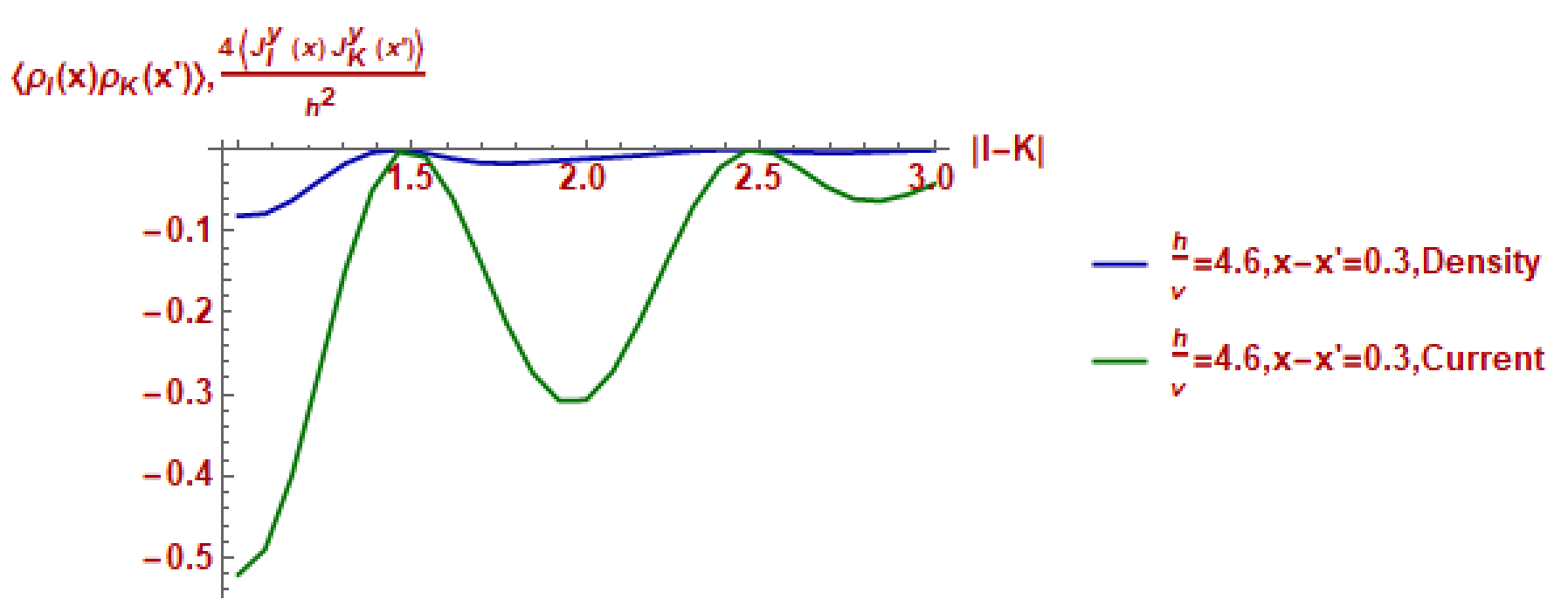}
\includegraphics[scale=0.8]{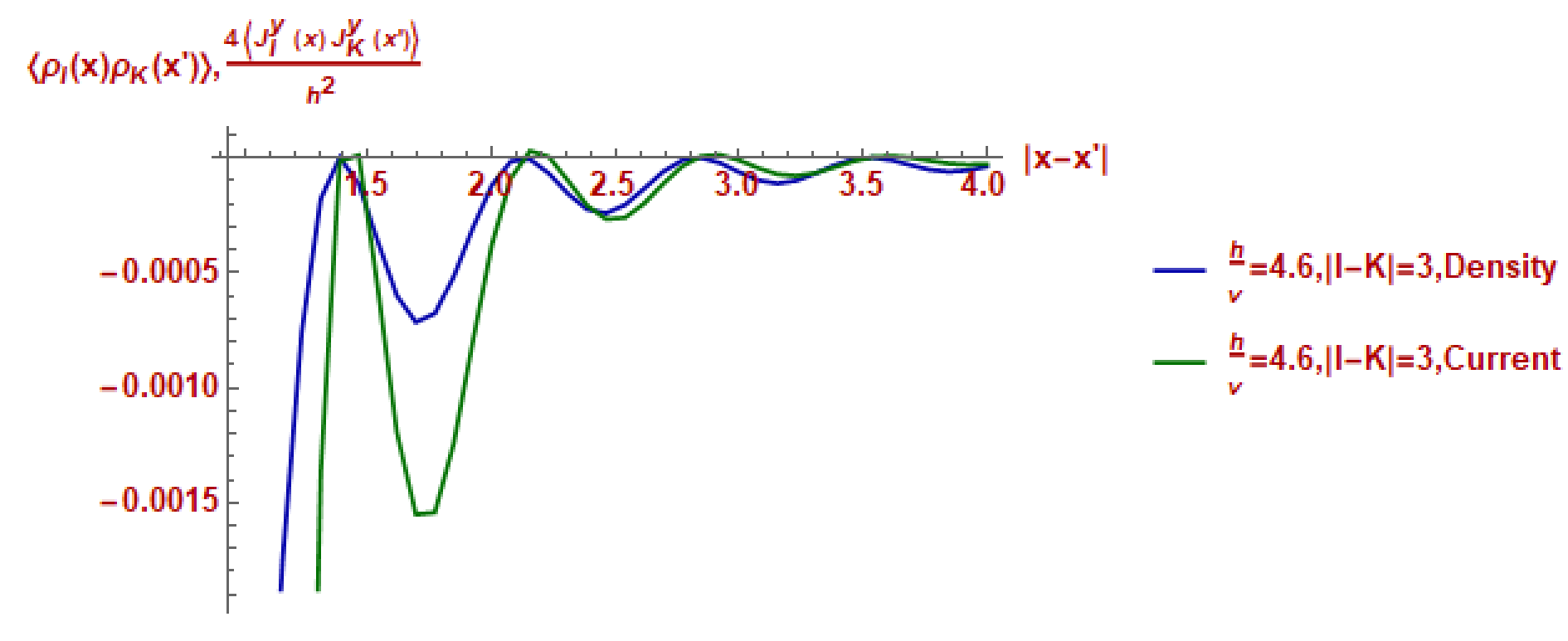}
\caption{Comparison of the $U(1)$ density and $y$ current two-point functions, normalized by $k$, as a function of $|I-J|$ at fixed $|x - x'|$ (top) and $|x - x'|$ at fixed $|I - K|$ (bottom).}
\label{currentcomparison}
\end{figure}
We are interested in computing the two-point correlation function,
\begin{align}
\langle J^y_I(x) J^y_K(x') \rangle_{k}.
\end{align}
As before, the $k>1$ correlation functions are proportional to the $k=1$ result:
\begin{align}
\langle J^y_I(x) J^y_K(x') \rangle_{k} = k \langle J^y_I(x) J^y_K(x') \rangle_{k = 1}.
\end{align}
$\langle J^y_I(x) J^y_K(x') \rangle_{k = 1}$ can be computed using the free fermion representation of the $k=1$ theory.
Mirroring the computation in the previous section, we find
\begin{align}
 \langle J^y_I(x) J^y_K(x') \rangle_{k = 1} & = - {h^2 \over 4 \pi^2}
\Tr \Big( M^\dagger(x) P^{I} \, M(x)  
\;  M^\dagger(x') P^{K} \, M(x') \Big)  \times {1 \over (x - x')^2}.
\end{align}
Comparisons of this $U(1)$ $y$-current two-point function---evaluated using the same method as in the previous section---and the density two-point function are shown in Fig.~\ref{currentcomparison} when $N=40$.

\section{Discussion}

\label{discussionsection}

In this paper, we proposed Wess-Zumino-Witten (WZW) theories with $U(N)$ symmetry at level $k\geq 1$ in two spacetime dimensions to be spinless (non-)Fermi liquid metals in two spatial dimensions.
The extra spatial dimension emerges from the $U(N)$ flavor symmetry of the WZW model.
The parameter $N \rightarrow \infty$ counts the number of points on the Fermi surface.
The $k=1$ theory is equivalent to Balents and Fisher's free 2d chiral metal.
The $k>1$ theories are interacting generalizations, in which the symmetries of the $k=1$ theory are maintained.
This construction provides a simple illustration of the ersatz Fermi liquid proposal of Else, Thorgren, and Senthil \cite{PhysRevX.11.021005}, specifically, that there are non-Fermi liquids that maintain the same symmetry as the Fermi liquid.
Here, it is the $U(N)$ symmetry and translation invariance (along the $x$ direction) of the free 2d chiral metal that are preserved at nonzero interaction $k>1$.
(The $U(N)$ symmetry includes translation along the $y$ direction.)
We computed the two-point function of the $U(1)$ number density and current operators for general $k \geq 1$.
The level $k$ results in an overall multiplicative rescaling of the amplitude of decay of these local correlation functions \footnote{The absence of non-Fermi liquid behavior in the $U(1)$ density and current two-point functions is similar to what occurs in the spinon-gauge problem \cite{KimFurusakiWenLee1994}.}.

There are a number of directions of future research.

$\bullet$ The most straightforward generalization is to non-chiral metals with an open Fermi surface, which would arise from arrays of non-chiral Luttinger liquids.
It would be interesting to consider the Lifshitz transition to a closed Fermi surface and/or potential instabilities of the non-chiral theories.
Another possibility is to gauge some of the $U(N)$ symmetry. 
This symmetry is anomalous in the chiral metals considered here; the non-chiral generalizations admit anomaly-free subgroups (see, e.g., \cite{PhysRevD.47.4546}).

$\bullet$ The nonlocal $SU(N)$ symmetry can be maintained in the presence of nonuniform hopping and/or random scalar potential quenched disorder \footnote{Nonuniform hopping $h_I$ and/or scalar potential disorder $V_I(x)$ can be gauged away using $M(x) = {\cal T}_x \exp \big( - i \int^x_{x_0} dy \big( \sum_{m = 1}^N V_m(y) \mathbb{I}^{(m)} -  {h_m \over v} X^{(m, m+1)} \big)\big)$, where ${\cal T}_x$ is an $x$-ordering operator, $x_0$ is an arbitrary basepoint, and $\mathbb{I}_{I J}^{(m)} = \delta_{m, I} \delta_{m, J}$.}. 
(This fact is essential to the theory of neutral modes in quantum Hall edge-state theories \cite{PhysRevLett.65.1502, KFP, PhysRevB.51.13449}.)
This may be useful for the study of $k>1$ models when translation invariance is broken.
The $k=1$ theory avoids localization because of its nonzero chirality \cite{Balents96, Balents97}.

$\bullet$ WZW theories admit free field representations \cite{Fuchs:1986ey}. 
This representation corresponds to free fermions when $k=1$.
The free field formulation may allow for the study of analogs of the ``single-particle" two-point function.
The decay of these correlation functions should exhibit exponents that depend on $k$, in contrast to the $U(1)$ density and current two-point functions we considered.

$\bullet$ It is important to find a microscopic realization for how $k$ may be varied, say, from $k=1$ to $k=2$. We reviewed one construction in \S\ref{WZWsection}, which relied on nonlocal interactions from the point of view of the underlying fermions.
Whether this deformation can be achieved via coupling to a local field is an open question.


\acknowledgments 

We thank Mike Hermele, Shamit Kachru, Chetan Nayak, Sri Raghu, Kirill Shtengel, and Shan-Wen Tsai for useful conversations and correspondence. 
We also thank Pak Kau Lim and Jeffrey Teo for collaboration on related matters.
This material is based upon work supported by the U.S. Department of Energy, Office of Science, Office of Basic Energy Sciences under Award No. DE-SC0020007.

\bibliography{bigbib}

\end{document}